\documentclass{article}

\begin{document}

\title{A short remark on the velocities of neutrinos }
\author{Elmir Dermendjiev \\
\ \textquotedblleft Mladost-2\textquotedblright , block 224, 1799-Sofia,
Bulgaria}
\maketitle

\begin{abstract}
An attempt is made to estimate the velocity $V_{\nu }$ of electron
antineutrino by using a theory of nuclear reactions \cite{1}. To
estimate
it, the experimental values of absorption cross sections $\mathbf{\sigma }%
_{a}(\widetilde{\nu }_{e})$ and $\mathbf{\sigma }_{a}(n)$ for electron
antineutrinos and neutrons by nuclei are compared. Based on the approach
proposed below, one could assume that the maximum velocity at weak
interactions $c_{w}>c,$ which is inconsistent with the theory of relativity
(TR).

antineutrino; Theory of relativity.
\end{abstract}

The quantum theory of nuclear reactions \cite{1} is based on the
assumption that all elementary particles have wave properties. Each
particle with mass $M$ and velocity $V$ has the wave length

\begin{equation}
\lambda =\frac{h }{MV}
\end{equation}

For neutral particles, such as a neutron, the theory \cite{1} $\
$gives simple estimation of the maximum value of absorption cross
section $\sigma
_{a}$, when a neutron with angular momentum $l=0$ is absorbed by a nucleus:%
\begin{equation}
\sigma _{a}(n)\approx \pi (R+\frac{\lambda _{n}}{2\pi})^{2}\frac{4k_{n}K_{n}}{%
(k_{n}+K_{n})^{2}}
\end{equation}

Here $K_{n}$ and $k_{n}$ are the wave vectors inside and outside of the
target nucleus with a radius $R$. For slow neutrons $\lambda _{n}>R$. One
can assume that at not very high kinetic energies of neutrons roughly $%
K_{n}\approx k_{n}$ \cite{1}. Then one finds that the value of%
\begin{equation}
\sigma _{a}(n)\approx  \frac{\lambda _{n}^{2}}{4\pi}
\end{equation}

The electron antineutrino $\widetilde{\nu }_{e}$, like a neutron, is a
neutral particle with spin $s=%
{\frac12}%
$. However, there is an essential difference between the physical
properties of these two particles. Electron antineutrino appears at
weak interactions. Contrary to the neutrino, a neutron is a massive
strong interacting particle. Nevertheless, based on the wave
properties of both particles one can assume that the interaction of
electron antineutrino with a nucleus could be formally considered in
the frames of quantum theory of nuclear reactions \cite{1}. Then one
could apply similar equations to estimate the
absorption cross section $\mathbf{\sigma }_{a}(\widetilde{\nu }_{e})$ of $%
\widetilde{\nu }_{e}$ by a target nucleus.

Below, it is assumed that the mass of electron antineutrino $M_{\widetilde{%
\nu }}>0$. Then, to estimate the value of $\mathbf{\sigma }_{a}(\widetilde{%
\nu }_{e})$ one can write the following equation that is similar to Eq.(2):

\begin{equation}
\sigma _{a}(\widetilde{\nu }_{e})\approx \pi (R+\frac{\lambda _{\widetilde{\nu }%
_{e}}}{2\pi})^{2}\frac{4k_{\widetilde{\nu }_{e}}K_{\widetilde{\nu }_{e}}}{(k_{%
\widetilde{\nu }_{e}}+K_{\widetilde{\nu }_{e}})^{2}}
\end{equation}

Since the mass $M_{\widetilde{\nu }}$ of $\widetilde{\nu }_{e}$ is expected
to be very small, i.e. $M_{\widetilde{\nu }}<<M_{n}$ , the binding energy
due to the absorption of $\widetilde{\nu }_{e}$ can be neglected and
approximately $K_{\widetilde{\nu }_{e}}\approx k_{\widetilde{\nu }_{e}}$.
The kinetic energy $E_{\widetilde{\nu }_{e}}$ of the electron antineutrino
that appears at $\beta ^{-}$ -- decay of nuclei usually does not exceed $%
0,5-1\ MeV$. Assuming the same value of kinetic energy of the $\widetilde{%
\nu }_{e}$ that interacts with a nucleus, one can suppose that $\lambda _{%
\widetilde{\nu }_{e}}>R$. Then, one gets simple relationship that can be
used for estimation of the value of $\sigma _{a}(\widetilde{\nu }_{e})$:

\begin{equation}
\sigma _{a}(\widetilde{\nu }_{e})\approx  \frac{\lambda _{\widetilde{\nu }%
_{e}}^{2}}{4\pi}
\end{equation}

Two approaches of estimation of values of $V_{\widetilde{\nu}_{e}}$
 are considered below.

One can calculate the value of $\lambda _{\widetilde{\nu }_{e}}$,
which is proportional to $\sqrt{\sigma _{a}(\widetilde{\nu }_{e})}$
and compare it with the measured value of $\sigma
_{a}(\widetilde{\nu }_{e})$ that was found to be of
$(10^{-43}-10^{-45})cm^{2}\ $\cite{2,3}. Assuming that
$E_{\widetilde{\nu }_{e}}\approx 1MeV$ and taking into account the
relativistic increase of the mass $M_{\widetilde{\nu }_{e}}$ at $V_{%
\widetilde{\nu }_{e}}\rightarrow c$ one gets a simple relationship for $%
\lambda _{\widetilde{\nu }_{e}}$:%
\begin{equation}
\lambda _{\widetilde{\nu }_{e}} = \frac{hc}{E_{\widetilde{\nu}_{e}}}
\end{equation}

Thus, the estimated values of $\lambda _{\widetilde{\nu }_{e}}$ and of $%
\sigma _{a}(\widetilde{\nu }_{e})$ are $\approx 10^{-6}cm$ and
$\approx 1,5.10^{-22}cm^{2}$ respectively. The discrepancy between
measured and estimated values of cross sections is of about $21-23$
orders of magnitude.
To reduce such huge difference one must assume that $\widetilde{\nu }_{e}>>c$%
, which is inconsistent with the TR. Also, to satisfy the value of $E_{%
\widetilde{\nu }_{e}}=1MeV$, one should assume that the mass $M_{0^{_{%
\widetilde{\nu }_{e}}}}$ at $V_{\widetilde{\nu }_{e}}=0$ is much less than $%
1eV$.

The ratio $r(\sigma )$ of both absorption cross sections from Eq.(3) and
Eq.(5) also allows the velocity of $V_{\widetilde{\nu }_{e}}$ to be
estimated:%
\begin{equation}
r(\sigma )=\frac{\sigma _{a}(\widetilde{\nu }_{e})}{\sigma _{a}(n)}=\frac{%
M_{n}^{2}V_{n}^{2}}{M_{\widetilde{\nu }_{e}}^{2}V_{\widetilde{\nu }_{e}}^{2}}
\end{equation}

Here, the values of $\sigma _{a}(\widetilde{\nu }_{e})\approx 10^{-43}cm^{2}$
\cite{2}, $\sigma _{a}(n)\approx 10^{-24}cm^{2},E_{\widetilde{\nu }%
_{e}}\approx 1MeV,\ M_{n}\approx 939MeV$ and the neutron velocity $%
V_{n}\approx 10^{8}cm/s$ are used. It is assumed that at $V_{\widetilde{\nu }%
_{e}}\rightarrow c$ the value of $M_{\widetilde{\nu }_{e}}=\frac
{E_{\widetilde{\nu }_{e}}}{c^2}$. The result is surprising: the value of $V_{%
\widetilde{\nu }_{e}}$ is estimated to be of about $10^{30}cm/s$. That value
is much larger than the value of $c$ given by the TR.

It is interesting to note that the ratio $t_{\widetilde{\nu }%
_{e}}/t_{nucl}\approx 10^{-19}.$ Here, $t_{\widetilde{\nu }_{e}}\approx
10^{-42}s$ is the antineutrino's flight-time through a nucleus and $t_{nucl}$
$\approx 10^{-23}s$ is the characteristic nuclear time. The value of the
ratio $r(\sigma )$ is also of $\approx 10^{-19}$. These values could be
associated with the \textquotedblleft $1/V$\textquotedblright - law for low
energy neutron cross sections. Indeed, the higher is the velocity of a
particle the less is an interaction time between that particle and the
target nucleus.

It is evident that so large estimated value of $V_{\widetilde{\nu
}_{e}}$ is inconsistent with the TR. As a matter of fact, following
the TR we assume that the values of V of elementary particles, such
as electrons, nucleons,
etc. cannot exceed the value of c. Moreover, many experimental data at $%
V\leq c$ confirm this grate theory. However, if the approach
discussed above is more or less correct, then one could assume that
at weak interactions the maximum permitted velocity $V_{w}>c$.
Unfortunately, up to now no measurements of the values of
$V_{\widetilde{\nu}_{e}}$ have been performed. On the other hand,
the problem of whether $c_{w}>c$, exceeds the frames of the TR. The
importance of this problem for physics, astronomy, cosmology, etc is
evident. A need of measurements of the values of
$V_{\widetilde{\nu}_{e}}$ is out of any doubts.

In conclusion, the author expresses his thanks to Dr.Angel Stanolov for
fruitful discussions of this short communication.

\end{document}